\documentclass[preprint,aps,showpacs,preprintnumbers,amsmath,amssymb]{revtex4}

\usepackage{graphicx,epsfig}% Include figure files
\usepackage{float}
\usepackage{dcolumn}% Align table columns on decimal point
\usepackage{bm}% bold math

\begin{document}

%\preprint{APS/123-QED}

\title{Effect of Nonlocal Spin-Transfer Torque on Current-Induced Magnetization Dynamics}
\author{Kyung-Jin Lee}
\email{kj_lee@korea.ac.kr}
\affiliation{Department of Materials
Science and Engineering, Korea University, Seoul 136-701, Korea}

\date{\today}% It is always \today, today,

\begin{abstract}
Using the self-consistent model, we present nonlocal spin-transfer
effects caused by the feedback between inhomogeneous magnetization
and spin-transfer torque on the current-induced magnetization
dynamics in nanomagnets. The nonlocal effects can substantially
improve the coherence time of precession in nanomagnets and thus
reduce the linewidth of power spectrum. This narrow linewidth
results from the nonlinear damping of spin-waves due to the nonlocal
spin torque which is inherent and thus should be considered in
future experiments.
\end{abstract}


\maketitle

%%%%%%%%%%%%%%%%%%%%%%%%%%%%%%%%%%%%%%%%%%%%%%%%%%%%%%%%%%%%%%%%%%%%%%%%%%%%%%
A spin-polarized current exerts a torque to a ferromagnet (FM) by
transferring the spin-angular momentum, i.e. spin-transfer
torque~\cite{Slon} (STT). STT generates a new class of magnetization
(\textbf{M}) dynamics in spin-valve structures~\cite{Myer, Kiselev},
potentially useful for applications in magnetic nonvolatile memories
and microwave oscillators.

In layered structures where the current flows perpendicular to the
plane, the direction and magnitude of STT at a point \textbf{r} is
decided by the spin accumulation $\bm{\mu}_s$ and associated spin
current $\mathbf{J}_s$ at the same point \textbf{r}. Initial
theories~\cite{Slon} assumed that the dependence of $\bm{\mu}_s$ on
\textbf{M} is local and thus essentially fixed by the local
\textbf{M} at the same point \textbf{r}. However, its dependence on
\textbf{M} is inherently nonlocal because of the 3-dimensional (3D)
spin diffusion~\cite{Polianski, Stiles, Brataas}. In other words,
when the conduction electron arrives at a point \textbf{r} on the
FM$|$normal metal (NM) interface, the reflected (transmitted)
electron takes the spin direction anti-parallel (parallel) to the
local \textbf{M} at the point \textbf{r}, diffuses along the
interface, and then transfers its spin-angular momentum to another
local \textbf{M} at a \textit{far away} point from the \textbf{r}.
That is, $\bm{\mu}_s$ at a point \textbf{r} is affected by all local
\textbf{M}'s at other points. The local assumption becomes really
invalid when \textbf{M} is inhomogeneous. Note that
micromagnetic~\cite{Miltat, LeeNat} and time-resolved imaging
studies~\cite{Acremann} have revealed excitations of incoherent
spin-waves and thus inhomogeneous \textbf{M} due to STT. In this
situation, the effect of $\bm{\mu}_s$ on \textbf{M} (=STT) and the
nonlocal effect of \textbf{M} on the $\bm{\mu}_s$ should be treated
on an equal footing. The conventional treatments, which ignore the
latter part, actually deal with only half of the relevant parts.
Therefore, the self-consistent feedback between inhomogenous
\textbf{M} and STT through the nonlocal effect should be considered.

The STT caused by the nonlocal effect can be named nonlocal
spin-transfer torque (NLST) since it allows a single FM with
inhomogeneous \textbf{M} to exert spin-transfer effects on itself.
Despite efforts to investigate NLST, the understanding of the
\textbf{M} dynamics affected by NLST remains elusive especially for
the spin-valve structure which is important from the viewpoints of
fundamental physics and applications. Previous theoretical studies
on NLST~\cite{Polianski, Stiles, Brataas} have addressed the
phenomenon in the perturbative regime of small spin-wave amplitudes
and thus could not investigate the dynamic modes for the current
exceeding the threshold for the onset of magnetic excitation.
Previous numerical studies~\cite{Adam} lacked the exact calculation
of 3D dynamic motion of $\bm{\mu}_s$ and focused only on the single
FM.

In this Letter, we have directly calculated 3D dynamic motion of
{$\bm{\mu}_s$} self-consistently coupled with the \textbf{M}
dynamics, which allows us to apply the model to both single FM and
spin-valve structures. This self-consistent treatment is essential
to correctly describe unique spin-wave modes caused by NLST and
explains two important experimental results: spin-wave excitations
in a single FM~\cite{Ozyilmaz} and narrower linewidths in
spin-valves than are expected within the assumption of homogeneous
\textbf{M}~\cite{Sankey}.

The equations of motion of \textbf{M} (Eq. (1)) and
{\boldmath$\mu_s$} (Eq. (2))~\cite{Polianski, Stiles, Brataas} are
self-consistently solved for FM and NM.
%%%%%%%%%%%%%%%%%%%%%%%%%%%%%%%%%%%%%%%%%%%%%%%%%%%%%%%%%%%%%%%%%%%%%%%%%%%%%%%%%%%%%
\begin{eqnarray}\label{LLG}
{\partial_t \mathbf{m}}= &-&\gamma_F (\mathbf{m} \times
\mathbf{H}_{eff}) + \alpha \mathbf{m} \times {\partial_t \mathbf{m}}  \\
&+& {\gamma_F /(M_s t_F)} \left[\mathbf{J}_s
|_{-{t_F/2}}-\mathbf{J}_s |_{+{t_F/2}}\right],  \nonumber \\
{\partial_t \bm{\mu}_s} &+& \nabla \cdot \mathbf{J}_s = -\gamma_{N}
(\bm{\mu}_s \times \mathbf{H}_{ext}) -
 {\bm{\mu}_s / \tau_{sf}}.
\end{eqnarray}
%%%%%%%%%%%%%%%%%%%%%%%%%%%%%%%%%%%%%%%%%%%%%%%%%%%%%%%%%%%%%%%%%%%%%%%%%%%%%%%%%%%%%
Here $\mathbf{m}$ is the unit vector of \textbf{M}, $\gamma_{F(N)}$
is the gyromagnetic ratio of FM (NM), $\mathbf{H}_{eff}$ is the
effective field including the magnetostatic, exchange, external
($\mathbf{H}_{ext}$), current-induced Oersted, and thermal
fluctuation fields, $\alpha$ is the intrinsic damping constant,
$M_s$ is the saturation magnetization, $t_F$ is the thickness of FM,
$\mathbf{J}_s=-D \nabla \bm{\mu}_s$ is the spin current, $D$ is the
diffusion coefficient, $\tau_{sf} = l_{sf}^2/D$ is the spin-flip
scattering time, and $l_{sf}$ is the spin-diffusion length. The
change of charge and spin current $J_e$ and $\mathbf{J}_s$ at the
interface of FM$|$NM are related to the potential drop over the
interface as~\cite{Circuit}
%%%%%%%%%%%%%%%%%%%%%%%%%%%%%%%%%%%%%%%%%%%%%%%%%%%%%%%%%%%%%%%%%%%%%%%%%%%%%%%%%%%%%
\begin{eqnarray}\label{BC1}
J_e &=&(G_\uparrow+G_\downarrow) \Delta \mu_e /e+
(G_\uparrow-G_\downarrow)\mathbf{m} \cdot (\Delta \bm{\mu_s}/e) \\
\mathbf{J}_s &=& (\hbar/2e^{2}) [Re(G_{\uparrow \downarrow})
\mathbf{m} \times (\mathbf{m} \times 2 \Delta \bm{\mu_s} \pm \hbar
\partial_t
\mathbf{m}), \nonumber \\
&-&\left((G_\uparrow+G_\downarrow) \mathbf{m} \cdot \Delta
\bm{\mu_s} - (G_\uparrow-G_\downarrow)\Delta \mu_e \right)
\mathbf{m}],
\end{eqnarray}
%%%%%%%%%%%%%%%%%%%%%%%%%%%%%%%%%%%%%%%%%%%%%%%%%%%%%%%%%%%%%%%%%%%%%%%%%%%%%%%%%%%%%
where $\mu_e$ is the electric potential, $\Delta \mu=\mu(\pm
t_F/2+0)-\mu(\pm t_F/2-0)$ is the potential drop over the interface,
$G_s$ (s=$\uparrow$ or $\downarrow$) is the spin-dependent
conductivity, $\beta(\gamma) =
(G_\uparrow-G_\downarrow)/(G_\uparrow+G_\downarrow)$ is the bulk
(interface) spin asymmetry, $G_{\uparrow\downarrow}$ is the mixing
conductivity. A small $Im(G_{\uparrow\downarrow})$ is
disregarded~\cite{Zwierzycki}. At the interface of FM$|$NM, $J_e$
and $\mathbf{J}_s \cdot \mathbf{m}$ are continuous under the
condition of $\bm{\mu}_s \times \mathbf{m}=0$ in FM. $\bm{\mu}_s$
and $\mathbf{m}$ are related through the Eqs. (2)-(4), and the
spin-version of the Ohm's law with the boundary conditions of
$\mu_e=-eV(0)$ and $\bm{\mu}_s=\bm{0}(\bm{0})$ at the far-right
(-left) end of the NM electrodes.

To validate the self-consistent model, we first carried out
simulations for the single FM,
Cu$_1(10)|$Co($t_{Co}$)$|$Cu$_2(52-t)$ (all in nm) where $t_{Co}$
varies from $2$ to $8 nm$, and compared modeling results to the
experimental ones in the Ref.~\cite{Ozyilmaz}. Since this structure
has no second FM, the conventional LLG-Slonczewski equation is not
applicable. Asymmetric Cu leads provide asymmetric $\bm{\mu}_s$ at
each side of the Co layer (Fig. 1(a)). $\bm{\overline{\mu}}$ at
interfaces ($= \bm{\mu}_s^{Cu_1|Co}+\bm{\mu}_s^{Co|Cu_2}$) is
negative when the electron flows from the thick to thin Cu layers,
corresponding to a negative current. This negative
$\bm{\overline{\mu}}$ provides negative NLST. Fig. 1(b) shows the
time evolution of averaged out-of-plane component of \textbf{M}
($\langle M_z \rangle$) at various negative currents when the
out-of-plane field $H$ is $2.5 T$. \textbf{M} initially saturates
along the out-of-plane direction, but cannot keep the saturation
state at negatively large currents even when $H$ is larger than the
out-of-plane demagnetization field $H_d$ ($\approx 1.6 T$) (Fig.
1(c)). When the current is turned on, a tiny in-plane component of
\textbf{M} is developed especially at the long edges where the
Oersted field is the largest. Interplay between this inhomogeneous
\textbf{M} and negative NLST excites spin-waves, resulting in the
rapid decrease of $\langle M_z \rangle$ within a few nanoseconds.

As in the experiment~\cite{Ozyilmaz}, we observed current-induced
excitations only at negative currents. The normalized modulus of the
magnetic moment $|M|$ is much smaller than $M_s$ at those bias
conditions (Fig. 1(d)), indicating excitation of large amplitude
incoherent spin-waves. Inset of Fig. 1(d) shows a snapshot of domain
pattern at $H=2.5 T$ and $I=-11mA$. Local \textbf{M}'s at edges are
mostly in the plane whereas those near the center of cell are in
vortex-like states caused by negative NLST which prefers
non-collinear configuration of local \textbf{M}'s. This
inhomogeneous \textbf{M} results in the reduction of the average
spin accumulation in the NM and thus the reduction of resistance of
the stack (not shown).

When $H>H_d$, the critical current $I_C$ for excitations linearly
depends on $H$ (Fig. 1(c) and (d)). As shown in Fig. 1(e), numerical
results of the slope ($=dI_C/dH$) are in better agreement with the
experimental ones than the theoretical ones (for the theoretical
$I_C$, see Eq. (10) in the Ref.~\cite{Polianski}). In the
experiment~\cite{Ozyilmaz}, the intercept of extrapolated boundary
at $I = 0$ is nearly zero for the sample with $t_{Co} = 8 nm$,
whereas the theoretical intercept is about $0.8 T (\approx H_d/2)$
for all thicknesses (inset of Fig. 1(e)). For $t_{Co} = 8 nm$, the
numerical intercept is $0.23 T$ and again in better agreement with
the experimental one. We attribute these better agreements to the
fact that the self-consistent model more realistically takes into
account the influence of the shape and finite size of nano-pillar on
the spin-wave mode.

Fig. 1(f) and insets show eigenmode analysis for the \textbf{M}
dynamics at $I = -11 mA$ and $H = 2.5 T$, corresponding to a
periodic oscillation of $\langle M_z \rangle$. As shown in Fig.
1(b), however, the $\langle M_z \rangle$ oscillation is in general
nonperiodic for most negative currents due to highly nonlinear
coupling among local \textbf{M}'s through the NLST. A rule of the
bias condition for a periodic oscillation may exist but we could not
find it because of a fixed step size of $I$ and $H$ in our
simulations. At this bias condition, the power spectrum shows two
peaks at $f_L (=75.3 GHz)$ and $2f_L$ where $f_L=\gamma_{Co}H/2
\pi$. The eigenmode images (insets) show that the precession region
with a higher power is localized at edges. Note that these
eigenmodes are unique features of the NLST and not expected in the
field-driven excitation~\cite{McMichael}.

This result demonstrates one crucial implication of the NLST,
namely, destabilization effect of negative NLST on local
\textbf{M}'s. Via the spin diffusion, the electrons backscattered
from the FM destabilize local \textbf{M}'s whereas the electrons
transmitted through the FM stabilize. The two effects always exist
simultaneously but one dominates the other because the NM electrodes
and thus $\bm{\mu}_s$ are not symmetric. Since the sign of
$\bm{\mu}_s$ is reversed by changing the current polarity, the
stabilizing effect is expected for a positive current, i.e. positive
NLST. In the single FM excitation, we observed almost macrospin
behaviors for positive NLST. In a spin-valve, however, different
types of spin-wave modes are expected for positive NLST because the
local STT is nonzero and thus generates incoherent
spin-waves~\cite{LeeNat}.

In the second study, we applied the self-consistent model to a
spin-valve structure, Cu(80)$|$Py(20)$|$Cu(6)$|$Py(2)$|$Cu(2)
(Py=Permalloy) experimentally studied by Sankey \textit{et
al.}~\cite{Sankey}. They reported a surprising result that the
current-induced dynamic modes can generate narrower linewidths at
low temperatures than those expected within the macrospin
assumption. To investigate the origin of this experimental finding,
we performed simulations with three different approaches: i)
macrospin model (MACRO), ii) conventional micromagnetic model
without considering NLST (CONV), and iii) self-consistent model
(SELF). Fig. 2(a), (b) and (c) show contours of spectral density of
$\langle M_x \rangle$ as a function of $I$ at $4K$ when the
effective field of $500 Oe$ is applied along the in-plane easy axis
$(//x)$. The positive current corresponds to the electron-flow from
Cu(2) to Cu(6), and thus positive NLST. MACRO shows the well-known
red- and blue-shift depending on $I$ (Fig. 2(a)). CONV shows only
red-shift up to a critical current ($I_C^{CONV} \approx 2 mA$, Fig.
2(b)). When $I > I_C^{CONV}$, \textbf{M} dynamics in CONV becomes
complicated due to excitations of incoherent spin-waves. As
indicated by an arrow, we observed secondary peaks with about half
the frequency of main peaks, corresponding to the precession of end
domains~\cite{LeeNat}. In SELF, we observed similar secondary peaks
indicating non-single domain state, but much clearer peak structures
than CONV up to about $2.4 mA$ which is larger than $I_C^{CONV}$
(Fig. 2(c)). It indicates that the positive NLST provides a more
periodic oscillation than that obtained in CONV.

Fig. 2(d) shows power spectra obtained in the three models. At a low
temperature ($T$), SELF shows the narrowest linewidth whereas CONV
produces the broadest one due to excitations of incoherent
spin-waves. We calculated the $T$ dependence of linewidth from
lorentzian fits (Fig. 2(e)). At low temperatures, SELF provides
narrower linewidths than MACRO, consistent with the experimental
observation~\cite{Sankey}. Therefore, the positive NLST indeed
results in a substantial improvement of the coherence time of
precession although \textbf{M} is not in the single domain state. It
indicates that it is possible to reduce the linewidth by properly
controlling the NLST. In MACRO, the linewidth monotonously increases
with $T$. On the other hand, in SELF, the linewidth linearly depends
on $T$ for $T < 50 K$ and more rapidly increases at higher
temperatures.

The narrower linewidths in SELF are caused by two nonlinear effects
of the positive NLST: an increase of the effective exchange
stiffness in short range and an increase of the damping of
incoherent spin-waves in long range. As a result, the positive NLST
provides an additional nonlinear spin-wave damping. For a
spin-torque nano-oscillator, the linewidth $\Delta \omega$ in the
low-temperature limit is given by~\cite{Slavin}
%%%%%%%%%%%%%%%%%%%%%%%%%%%%%%%%%%%%%%%%%%%%%%%%%%%%%%%%%%%%%%%%%%%%%%%%%%%%%%%%%%%%%
\begin{equation}\label{linewidth0}
{\Delta \omega = \Gamma_+(P_0) \left({{k_BT}/E_0}\right)
\left[1+\left(N/\Gamma_{eff} \right)^2 \right]}
\end{equation}
%%%%%%%%%%%%%%%%%%%%%%%%%%%%%%%%%%%%%%%%%%%%%%%%%%%%%%%%%%%%%%%%%%%%%%%%%%%%%%%%%%%%%
where $N=d\omega(P)/dP$ is the nonlinear frequency shift coefficient
obtained from $\omega(P)=\omega_0+NP$, $\omega_0$ is the
ferromagnetic resonance frequency at $I=0$, $P$ is the normalized
power, $\Gamma_{eff}=\sigma(I+QI_c)$ is the effective nonlinear
damping, $Q$ is a phenomenological coefficient characterizing the
nonlinear positive damping, and $I_c$ is the critical current for
the magnetic excitation (for details of other parameters, see
Ref.~\cite{Slavin} ).

Eq. (\ref{linewidth0}) predicts two important consequences of the
nonlinearity. First, the linewidth of an auto-oscillator with a
nonlinear frequency shift (i.e. $N \ne 0$) increases by the factor
$(1+(N/\Gamma_{eff})^2)$ from that of a linear oscillator (i.e.
$N=0$). Second, the linewidth of a nonlinear oscillator decreases
with increasing the nonlinearity of damping $Q$. It is because the
linewidth is determined by nonlinear properties of the system where
the normal linear damping is compensated by local STT. In this case,
an increase of the nonlinearity of damping can lead to a decrease of
the linewidth, known as the noise suppression due to nonlinear
feedback~\cite{RMP}.

Inset of Fig. 2(e) shows that $N$ is nonzero and almost identical
for the two models. Thus, the linewidth is wider than that expected
in a linear oscillator. Using Eq. (\ref{linewidth0}), we fit the
values of $Q$ from the calculated linewidths at $T = 10 K$ and
obtained $Q = 0.13$ in MACRO and $Q = 1.96$ in SELF. The fit value
$Q$ in SELF is consistent with the assumed values $(Q=1 \sim
3)$~\cite{Slavin} to explain experimental observations. Note that
the nonlinear theory referred here does not take into account the
spin transport and the $Q$ value has been used as a fitting
parameter without justification of its origin. In contrast, our
self-consistent treatment shows that the large $Q$ is mainly caused
by NLST. Thus, we conclude that the nonlinear spin-wave damping due
to NLST is responsible for narrower linewidths in SELF at low
temperatures. When $T$ is too high, the thermal random force
overcomes the nonlocal effect due to positive NLST and thus the
linewidth abruptly increases. For the opposite current polarity
(i.e. negative NLST), we observed an increase of the linewidth (not
shown).

Finally, we note that the magnitude of NLST is easily controlled by
modifying the asymmetry of layer structure like the conventional
local STT. The effect of NLST on the current-induced \textbf{M}
dynamics is determined by the ratio of NLST to local STT. Fig. 2(f)
shows the ratio at the parallel magnetic configuration as a function
of the thickness of Cu spacer $(t_{Cu})$ for the spin-valve
structure studied here. The ratio is about $0.1$ at $t_{Cu} = 6 nm$
which is the case of the Ref.~\cite{Sankey}. Note that the effect of
NLST on the current-induced \textbf{M} dynamics is considerable
although the ratio is only $0.1$. Furthermore, this ratio increases
with increasing $t_{Cu}$ as shown in Fig. 2(f). Therefore, NLST
should be considered in designing and interpreting future
experiments.

We thank B. Dieny, A. Vedyayev, M. D. Stiles, A. Brataas, A. Kent,
J. Z. Sun, I. N. Krivorotov, A. N. Slavin, J. -V. Kim, S. Zhang and
H. -W. Lee for fruitful discussions. This work was supported by KRF
(MOEHRD) (KRF-2006-311-D00102), KOSEF through the NRL Program (No.
M10600000198-06J0000-19810), KISTI under the Strategic
Supercomputing Support Program.

%($\dagger$) Corresponding email: kj\_lee@korea.ac.kr.
%%%%%%%%%%%%%%%%%%%%%%%%%%%%%%%%%%%%%%%%%%%%%%%%%%%%%%%%%%%%%%%%%%%%%%%%%%%%%%%%%%%%%
%\\newpage %Just because of unusual number of tables stacked at end
%\\bibliography{apssamp}% Produces the bibliography via BibTeX.

\begin{thebibliography}{99}
%\bibitem[{$^\dagger$}]{} Corrresponding e-mail address: kj\_lee@korea.ac.kr
%\bibitem[$\dagger$]{Sejoong}  Present Address: Department of Physics,
%Massachusetts Institute of Technology, Cambridge, MA 02139, USA.

\bibitem{Slon} J. C. Slonczewski, J. Magn. Magn. Mater. \textbf{159}, L1 (1996); L. Berger, Phys. Rev. B \textbf{54}, 9353 (1996).

\bibitem{Myer} E. B. Myers \textit{et al.}, Science \textbf{285}, 867 (1999).

\bibitem{Kiselev} S. I. Kiselev \textit{et al.}, Nature (London) \textbf{425}, 380 (2003).

\bibitem{Polianski} M. L. Polianski and P. W. Brouwer, Phys. Rev. Lett. \textbf{92}, 026602 (2004).

\bibitem{Stiles} M. D. Stiles, J. Xiao, and A. Zangwill, Phys. Rev. B \textbf{69}, 054408 (2004).

\bibitem{Brataas} A. Brataas, Y. Tserkovnyak, and G. E. W. Bauer, Phys. Rev. B \textbf{73}, 014408 (2006).

\bibitem{Miltat} J. Miltat, G. Albuquerque, A. Thiaville, and C. Vouille, J. Appl. Phys. \textbf{89}, 6982
(2001); K.-J. Lee and B. Dieny, Appl. Phys. Lett. \textbf{88},
132506 (2006).

\bibitem{LeeNat} K.-J. Lee \textit{et al.}, Nat. Mater. \textbf{3}, 877 (2004).

\bibitem{Acremann} Y. Acremann \textit{et al.}, Phys. Rev. Lett. \textbf{96}, 217202
(2006); J. P. Strachan \textit{et al.}, Phys. Rev. Lett.
\textbf{100}, 247201 (2008).

\bibitem{Adam} S. Adam, M. L. Polianski, and P. W. Brouwer, Phys. Rev. B \textbf{73}, 024425
(2006); M. A. Hoefer, T. J. Silva, and M. D. Stiles, Phys. Rev. B
\textbf{77}, 144401 (2008).

\bibitem{Ozyilmaz} B. \"{O}zyilmaz \textit{et al.}, Phys. Rev. Lett. \textbf{93}, 176604 (2004).

\bibitem{Sankey} J. C. Sankey \textit{et al.}, Phys. Rev. B \textbf{72}, 224427 (2005).

\bibitem{Circuit} A. Brataas,  Y. V. Nazarov, G. E. W. Bauer, Phys. Rev.
Lett. \textbf{84}, 2481 (2000).

\bibitem{Zwierzycki} M. Zwierzycki \textit{et al.}, Phys. Rev. B \textbf{71}, 064420 (2005).

\bibitem{Bass} J. Bass and W. P. Pratt Jr., J. Magn. Magn. Mater. \textbf{200}, 274 (1999).

\bibitem{McMichael} R. D. McMichael and M. D. Stiles, J. Appl. Phys. \textbf{97}, 10J901 (2005).

\bibitem{Ilya} I. N. Krivorotov \textit{et al.}, Science \textbf{307}, 228 (2005).

\bibitem{Slavin} A. N. Slavin and K. Pavel, IEEE Trans. Magn. \textbf{41}, 1264
(2005); J. -V. Kim, V. Tiberkevich, and A. N. Slavin, Phys. Rev.
Lett. \textbf{100}, 017207 (2008); J. -V. Kim \textit{et al.}, Phys.
Rev. Lett. \textbf{100}, 167201 (2008).

\bibitem{RMP} J. Bechhoefer, Rev. Mod. Phys. \textbf{77}, 783 (2005).

\end{thebibliography}

%%%%%%%%%%%%%%%%%%%%%%%%%%%%%%%%%%%%%%%%%%%%%%%%%%%%%%%%%%%%%%%%%%%%%%%%%%%%%%%%%%%%%
\newpage

%%%%%%%%%%%%%%%%%%%%%%%%%%%%%%%%%%%%%%%%%%%%%%%%%%%%%%%%%%%%%%%%%%%%%%%%%%%%%%%%%%%%%
\begin{figure}[ttbp]
\begin{center}
\psfig{file=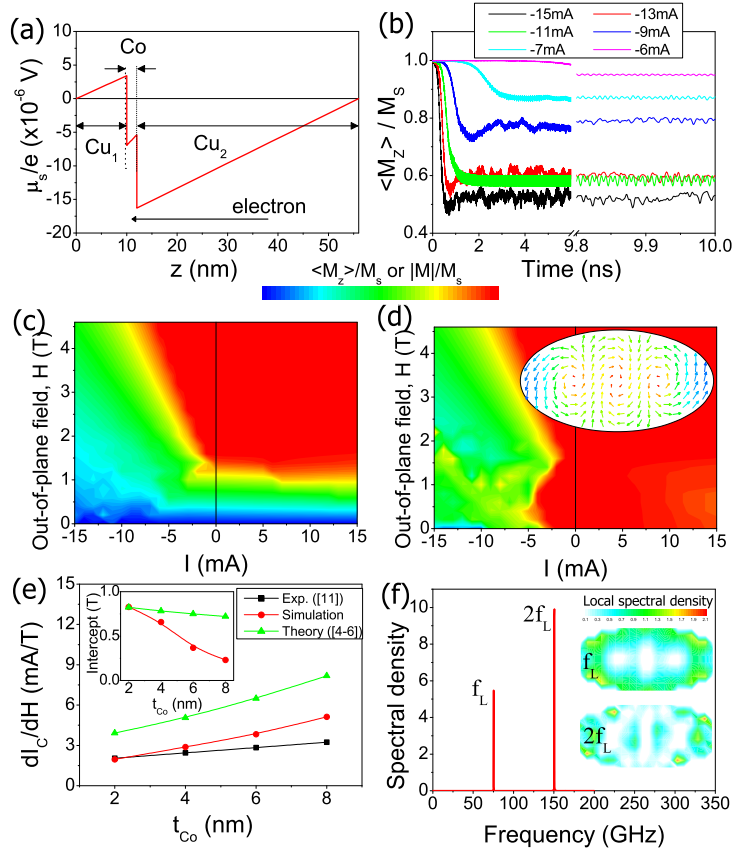,width=0.6\columnwidth} \caption{\label{fg1}
(Color online) Current-induced excitation of single ferromagnet
($t_{Co}=2nm$). (a) Spin accumulation pattern. (b) Time evolution of
$\langle M_z \rangle$ at $H = 2.5 T$ and various negative currents.
(c) and (d) show contours of $\langle M_z \rangle$ and $|M|$ as a
function of $H$ and $I$, respectively. Inset of (d) shows a domain
pattern obtained at $H=2.5 T$ and $I=-11mA$. (e) Slope of the
critical boundary ($=dI_C/dH$) as a function of $t_{Co}$. Inset of
(e) shows the intercept of the extrapolated boundary as a function
of $t_{Co}$. (f) Power spectrum at $H=2.5 T$ and $I=-11mA$. Insets
of (f) show eigenmode images for the two peak frequencies. Model
parameters: Elliptical shaped pillar with $60 \times 30 nm^2$,
$M_s=1420emu/cm^3$, the exchange stiffness constant $A_{ex}=2 \times
10^{-6} erg/cm$, $\alpha =0.01$, the unit cell size$=3nm$, and the
discretization thickness of Cu layer varies depending on the total
thickness and is not larger than $5 nm$. For Cu and Co, the spin
transport parameters~\cite{Bass} are bulk resistivity $\rho (\mu
\Omega cm)$=0.6 and 7.5, $\beta$=0 and 0.46, $l_{sf} (nm)$=450 and
59, and $D (\times 10^{15} nm^2s^{-1})$=41 and 1.7. For the
interface Co$|$Cu, the parameters are interfacial resistance $AR^*
(m \Omega \mu m^2)$=0.51, $\gamma$=0.77, interfacial spin memory
loss $\delta$=0.25, and $Re(G_{\uparrow \downarrow}) (\times 10^{10}
\Omega^{-1}cm^{-2})$=5.5.}
\end{center}
\end{figure}
%%%%%%%%%%%%%%%%%%%%%%%%%%%%%%%%%%%%%%%%%%%%%%%%%%%%%%%%%%%%%%%%%%%%%%%%%%%%%%%%%%%%%%%%

%%%%%%%%%%%%%%%%%%%%%%%%%%%%%%%%%%%%%%%%%%%%%%%%%%%%%%%%%%%%%%%%%%%%%%%%%%%%%%%%%%%%%
\begin{figure}[ttbp]
\begin{center}
\psfig{file=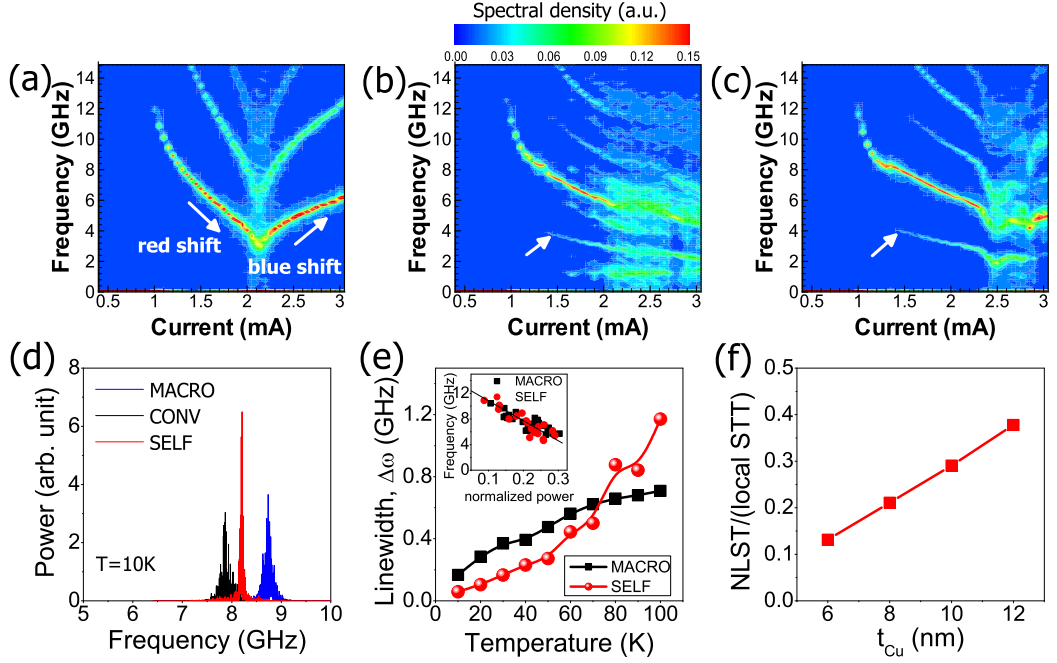,width=0.7\columnwidth} \caption{\label{fg2}
(Color online) Current-induced excitation of spin-valve. Contours of
spectral density of $\langle M_x \rangle$ at $T=4K$ obtained in (a)
MACRO, (b) CONV, and (c) SELF. (d) Comparison of power spectra
obtained in the three models at $T=10 K$. (e) Linewidth as a
function of the temperature. Inset of (e) shows the frequency versus
the power normalized by $|M|$. (f) The ratio of NLST to local STT as
a function of the thickness of Cu spacer ($t_{Cu}$). For (d) and
(e), we chose different bias conditions for each model to yield a
similar precession angle since the linewidth is proportional to the
precession angle~\cite{Kiselev}. However, the main finding is not
altered even when we choose the same bias condition for all models.
Model parameters: Elliptical shaped nanopillar with $120 \times 60
nm^2$, $M_s=645 emu/cm^3$~\cite{Ilya}, $A_{ex}=1.3 \times 10^{-6}
erg/cm$, $\alpha=0.025$~\cite{Ilya}, and the unit cell size=$5nm$.
For Py, the spin transport parameters are $\rho (\mu \Omega
cm)=25.5$, $\beta=0.7$, $l_{sf} (nm)=5.5$, and $D (\times 10^{15}
nm^2s^{-1})=1.7$. For the interface Py$|$Cu, the parameters are
$AR^* (m \Omega \mu m^2)$=0.97, $\gamma$=0.77, $\delta$=0.25, and
$Re(G_{\uparrow \downarrow}) (\times 10^{10}
\Omega^{-1}cm^{-2})$=6.0. Parameters of Py were provided by Cornell
group. $Re(G_{\uparrow \downarrow})$ of Py$|$Cu was determined to
mimic the critical current in the Ref.~\cite{Sankey}. The pinned
layer \textbf{M} is fixed along the in-plane easy axis (no stray
field from it).}
\end{center}
\end{figure}
%%%%%%%%%%%%%%%%%%%%%%%%%%%%%%%%%%%%%%%%%%%%%%%%%%%%%%%%%%%%%%%%%%%%%%%%%%%%%%%%%%%%%

\end{document}